# A Novel Distributed Database Architectural Model for Mobile Cloud Computing

Somenath Chakraborty[1], Dia Ali[1], Beddhu Murali[1]

[1] The University of Southern Mississippi,
Hattiesburg, Mississippi, USA
{Somenath.Chakraborty, Dia.Ali, Beddhu.Murali}@usm.edu

**Abstract.** Cloud computing is the way by which we connect to servers, large systems into a distributed secure manner without worrying about local memory limits. Here this paper we proposed a Novel Distributed Database Architectural Model for Mobile Cloud Computing (NDDAMMCC). Due to the exponential growth of wireless technologies and internet which are following Nielsen's Law of Internet Bandwidth, we are in the new era of cloud computing. In the recent technological era, smart mobile devices play a big role in all sort of day-by-day human needs. The applicability is so huge that the number of apps install on a mobile system becomes a hazard due to local memory limitations for mobile phone users and demands an alternative approach to solve this local memory problems. Mobile Cloud Computing (MCC) is the ultimate solution to this issue and our model presents a promising path in this new kind of cloud computing technology.

## 1   Introduction

The cloud computing is the way by which we connect servers, large systems into a distributed secure manner without worrying about local memory limits. Initially we only focused on desktop computing applications to run on cloud systems, but in the recent time smartphones becomes so powerful and evident part of our daily life, that we are now capable of performing any kind of computing applications by using only our smartphones. But smartphones faces new challenges specially due to the memory usage, processing infrastructure shortage due to the local smartphone RAMs and other mobile memory management limitations. Here this paper mainly focused on presenting distributed database architectural model for mobile cloud computing. With the advancement of Internet speed, Internet bandwidth, efficient dynamic channel allocation for wireless communication[5], Internet of Things(IoT) revolution, and the technological development of smartphone, we have much flexibility to develop mobile architectural model that can run on the cloud using Mobile Cloud

Computing(MCC) applications. MCC deployments mainly categorized into three classes of applications. The following Table 1 shows the listings.

**Table 1.** Mobile Cloud Computing Applications.

| Class | Usages and Properties | Examples |
| --- | --- | --- |
| Private | Very Secure Applications, All the Hardware not shared and own by the owner of the Private Cloud. | Mobile bank applications, Server Login applications etc. |
| Public | Anybody can freely access this kind of applications, it is shared on open internet so anybody can access this services, less Secure, hardware and software not own by the owner of Public cloud. It provides cost efficiency and promotes agile methodologies. | Any kind of free applications which are deployed on open internet. |
| Hybrid | Integration of Private and Public both layers. | Any MCC applications or CloudOps which consist of free and Purchased features of content and services, like credit check systems. |

## 2  Literature Review

According to NIST SP 800-145 [1,3], any cloud computing architectural model should support the following characteristics, on-demand self-services, ubiquitous network access, resource pooling, rapid elasticity, pay-per use.
They[1], also represent the standardization of three service model of Cloud Computing. These are Software as a Service(SaaS) model, a third party company or business, hosts application software for the end user, other components, like hardware, are maintained by the provider company. In Infrastructure as a Service (IaaS) Model, A third-party provider company hosts the software, hardware, services, and other aspects of the system. An organization that wants to maximize costs in the company and needs the scaling would choose this cloud. In a Platform as a Service (PaaS) model, the third-party company provides the deployment platforms for the end user. It supports cloud versions of application development, deployment of different services and hosting on cloud platform. The following Fig.1 portrays the visualization for these three types of cloud model. Zhang, Q., Cheng, L. & Boutaba, R. [2], addresses the issues and challenges faced by the cloud computing infrastructure development and recognize important research directions in this emerging field. Huang, D., & Wu, H. [4], describe in their book chapter, how different kinds of Mobile Cloud Computing Framework works and the basic implementation of their architecture, development challenges and basic properties.

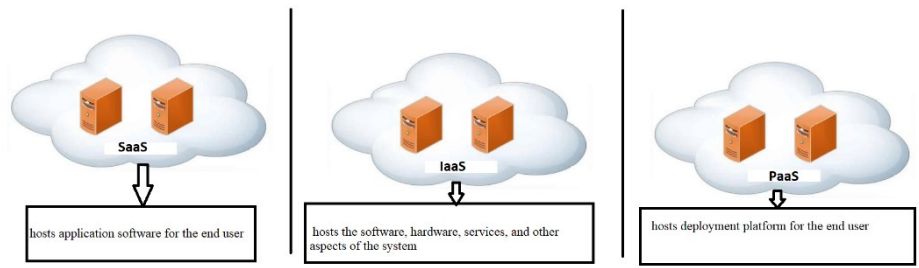

Fig. 1. Three Service Model of Cloud Computing

The Mobile Cloud Computing Forum [6], formulates the definition of MCC as a simplest way to support infrastructure, data storage and data processing should be performed from the outside of the mobile devices. The should be from the cloud and also broaden the services not only the smart phone users but a host of application supports to its subscribers as well. Aepona [7], shows the potential of mobile cloud computing with the high processing infrastructure support and huge fast processing memory support. Ruay Shiung Chang et. all., presents a comprehensive description of MCC issues, Challenges and provides the insights of the needs in present times where there is a demands created for different kinds of cloud applications.
Though all those paper provides insightful information and shows the pathways for new kinds of cloud support systems but many are mostly using traditional infrastructure and sometime lacks crucial efficiency and limitations in client demands. Here, in this paper, we proposed and MCC model which can leverages the advancement of distributed computing through sharing and processing data through distributed data sharing support. M.Rajendra Prasad, R. Lakshman Naik, V.Bapuji[10], outlines cloud computing research issues and implications and describes the comprehensive analysis of recent research in this domain.

## 3 Proposed Methodology

Our Novel Distributed Database Architectural Model for Mobile Cloud Computing (NDDAMMCC) is illustrated in the Fig. 2. The architecture supports the advanced distributed cloud supports which enable global agility and robustness to the overall architectural system. In this architectural model there are basically four paradigm of infrastructure embedded into a complete system model. It is shown in Fig. 3.

Different kinds of mobile devices gets the dynamic signals [5] in a portable moving environment where the Access point(AP's) establish and supports the wireless services by routing the signals coming from the Mobile base stations(BTS). A set of mobile networks connected through the distributed cloud infrastructure which are capable of providing all kinds of cloud services as well as distributed high speed memory and application processing support and other platforms support. These distributed clouds coupled in a independent architecture but for better security control these distributed system integrated with the central cloud systems. Through the mobile networks are shown in a integrated format in Fig. 2, but they are fully independent and their other kinds of mobile services do not rely on cloud infrastructure support. The distributed cloud also do not provide hardware or software support to the mobile service providers. So, individually any mobile user can subscribe these distributed cloud services though the internet and get all those enhanced experiences and facilities of mobile cloud computing(MCC).

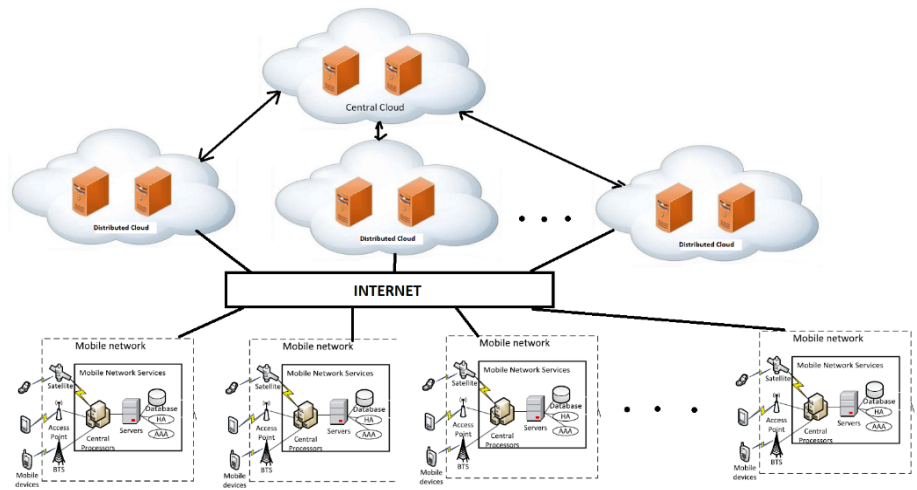

**Fig. 2.** A Novel Distributed Database Architectural Model for Mobile Cloud Computing (NDDAMMCC)

The Novel Distributed Database Architectural Model for Mobile Cloud Computing (NDDAMMCC) algorithm is presented in Algorithm 1.

Algorithm 1: The Novel Distributed Database Architectural Model for Mobile Cloud Computing (NDDAMMCC)

1. start
2. Input: hostList, ApplicationList  Output: Allocation of Distributed shared resource to ApplicationList.
3. DistributedCloud.Orderby.ResourceReq()
4. For each ResourceReq() in ApplicationList Do
5.    If ResourceReq() < DistributedCloud_Capacity()
6.      Allocate Demand resource
7.    Else ResourceReq = Satisfied

```
    8.        Do: Max_allocation >= Small_ApplicationList
    9.            Host_allocation with Accesspointsupport(mode=Dynamic)
   10.            Resource.Add  = Resource_Required( mode= dynamic)
   11.    Use Resource until ApplicationList finished
   12.    End
```

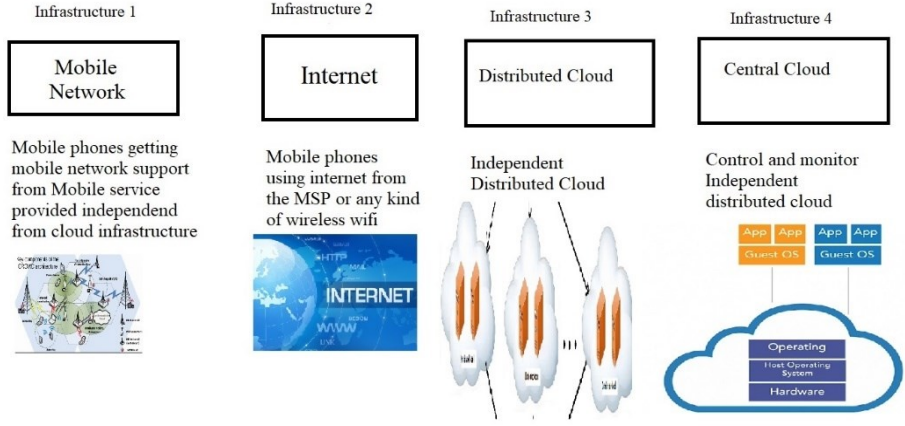

**Fig. 3.** Four Types of Infrastructure support for our Novel Distributed Database Architectural Model for Mobile Cloud Computing (NDDAMMCC)

## 4  Experimental Simulations

This distributed cloud are interconnected through distributed database support systems and monitor a distributed log which is auto-updated and protects all kinds of vulnerabilities and attracts from the outside world. We use Cloudsim Simulator [9], and the simulation results are presented with the following terminologies.

   A. Capacity: The total capacity of a host having np processing elements (PEs) is given by:

$$\text{Capacity} = \sum_{i=1}^{np} \frac{cap(i)}{np},$$

where cap(i) is the processing strength of individual elements.

A space-shared policy is applied for allocating VMs to hosts and a time-shared policy forms the basis for allocating task units to processing core within a VM. Hence, during a VM lifetime, all the tasks assigned to it are dynamically context switched during their life cycle. By using a time-shared policy, the estimated finish time of a Cloudlet managed by a VM is given by,

$$\text{eft}(p) = ct + \frac{rl}{capacity \times cores(p)}$$

where eft(p) is the estimated finish time, ct is the current simulation time, and cores(p) is the number of cores (PEs) required by the Cloudlet. In time-shared mode, multiple Cloudlets (task units) can simultaneously multi-task within a VM. In this case, the total processing capacity of Cloud host as,

$$\text{capacity} = \frac{\sum_{i=1}^{np} cap(i)}{max\left(\sum_{j=1}^{cloudlets} Cores(j), np\right)}$$

where cap(i) is the processing strength of individual elements.

**Table 2**: Analysis of results based on A Novel Distributed Database Architectural Model for Mobile Cloud Computing (NDDAMMCC)

| Distributed Cloud details(VMs) | Capacity(Dynamically varying) using space shared | Estimated finish time(in milisec) | Total processing capacity of Cloud host |
|---|---|---|---|
| 12 tasks in 3 VMs | 26 | 239.24 | 16 |
| 23 tasks in 8 VMs | 24 | 568.74 | 18 |
| 39 tasks in 12 VMs | 28 | 698.45 | 16 |

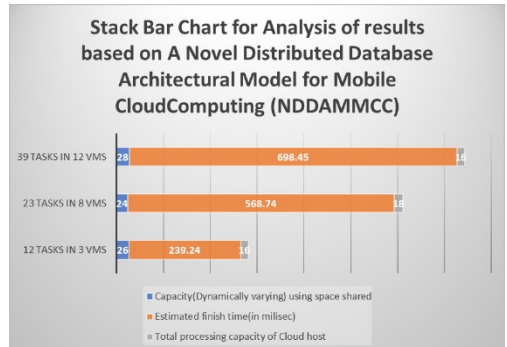

**Fig.4 Stack Bar Chart for Analysis of results based on A Novel Distributed Database Architectural Model for Mobile Cloud Computing (NDDAMMCC)**

## 5 Conclusion

The main advantages of this model is that if any system fails then also it provide uninterrupted services in a distributed manner. This proactive model prevents failures of the system as it is auto-managed in a dynamically space-resource by the central log from its distributed child nodes.

**Acknowledgments.** The authors acknowledge High Performance Computing (HPC) at "The University of Southern Mississippi" supported by the National Science Foundation(NFS), USA, under the Major Research Instrumentation (MRI) program via Grant # ACI 1626217.

# References

1. Simmon, E. (2018), Evaluation of Cloud Computing Services Based on NIST SP 800-145, Special Publication (NIST SP), National Institute of Standards and Technology, Gaithersburg, MD, [online], https://doi.org/10.6028/NIST.SP.500-322 (Accessed May 30, 2021).

2. Q. Zhang, L. Cheng and R. Boutaba, "Cloud computing: State-of-the-art and research challenges," Journal of Internet Services and Applications, vol. 1, no. 1, pp. 7–18, 2010.

3. Badger, L., D. Bernstein, R. Bohn, F. de Vaulx, M. Hogan, M. Iorga, J. Mao, J. Messina, K. Mills, E. Simmon, A. Sokol, J. Tong, F. Whiteside, and D. Leaf, US Government Cloud Computing Technology Roadmap - Volumes I and II, NIST SP 500-293. 2014, NIST. p. 40 (Vol I), 98 (Vol II).

4. Huang, D., & Wu, H. (2018). Mobile Cloud Computing Service Framework. Mobile Cloud Computing, 91–114. doi:10.1016/b978-0-12-809641-3.00006-5.

5. Somenath Chakraborty, Banani Saha and Samir K Bandyopadhyay. Dynamic Channel Allocation in IEEE 802.11 Networks. International Journal of Computer Applications 149(1):36-38, September 2016.

6. http://www.mobilecloudcomputingforum.com/.

7. White Paper. Mobile Cloud Computing Solution Brief. AEPONA, 2010.

8. Ruay Shiung Chang, Jerry Gao, Volker Gruhn, Jingsha He, George Roussos, and Wei-Tek Tsai, "Mobile Cloud Computing Research - Issues, Challenges and Needs", Proceedings of IEEE Seventh International Symposium on Service-Oriented System Engineering, San Francisco, USA, 2013.

9. Calheiros RN, Ranjan R, Beloglazov A, De Rose CA, Buyya R (2011). "CloudSim: a toolkit for modeling and simulation of cloud computing environments and evaluation of resource provisioning algorithms", Software: Practice and Experience. 41 (1): 23–50. doi:10.1002/spe.995.

10. M.Rajendra Prasad, R. Lakshman Naik, V.Bapuji,"Cloud computing: Research issues and implications",International Journal of Cloud Computing and Services Science (IJ-CLOSER) Vol.2, No.2, April 2013, pp. 133~139 ,ISSN: 2089-3337.

11. Chakraborty, Somenath, and Samir K. Bandyopadhyay. "Scene text detection using modified histogram of oriented gradient approach." IJAR 2, no. 7 (2016): 795-798.

12. Chakraborty, Somenath. "Category Identification Technique by a Semantic Feature Generation Algorithm." In Deep Learning for Internet of Things Infrastructure, pp. 129-144. CRC Press, 2021.

13. Chakraborty, Somenath, and Chaoyang Zhang. "Survival prediction model of renal transplantation using deep neural network." In 2020 IEEE 1st International Conference for Convergence in Engineering (ICCE), pp. 180-183. IEEE, 2020.

14. Chakraborty, Somenath, and Beddhu Murali. "Investigate the Correlation of Breast Cancer Dataset using Different Clustering Technique." arXiv preprint arXiv:2109.01538 (2021).

15. Chakraborty, Somenath, Beddhu Murali, and Amal K. Mitra. "An Efficient Deep Learning Model to Detect COVID-19 Using Chest X-ray Images." International Journal of Environmental Research and Public Health 19, no. 4 (2022): 2013.

16. Chakraborty, Somenath, and Beddhu Murali. "A Novel Medical Prognosis System for Breast Cancer." In Proceedings of International Conference on Advanced Computing Applications, pp. 403-413. Springer, Singapore, 2022.

17. Sanyal M., Chakraborty S. (2022) Enhanced Multigradient Dilution Preparation. In: Das K.N., Das D., Ray A.K., Suganthan P.N. (eds) Proceedings of the International Conference on Computational Intelligence and Sustainable Technologies. Algorithms for Intelligent Systems. Springer, Singapore. https://doi.org/10.1007/978-981-16-6893-7_46.

18. Chakraborty, Somenath. "A Local Outlier Mining Algorithm Based on Region Segmentation." 한국컴퓨터게임학회논문지 34, no. 3 (2021): 133-141.